# Single-cycle, 643-mW average power THz source based on tilted pulse front in lithium niobate

**Tim Vogel***, **Samira Mansourzadeh**, **Clara J. Saraceno**

*Photonics and Ultrafast Laser Science (PULS), Ruhr-University Bochum, 44801 Bochum, Germany*
*\*Tim.Vogel-u81@ruhr-uni-bochum.de*

We present, to the best of our knowledge, the highest average power from a laser-driven single-cycle THz source demonstrated so far, using optical rectification in the titled pulse-front geometry in cryogenically cooled lithium niobate, pumped by a commercially available 500 W ultrafast thin-disk Yb-amplifier. We study repetition rate dependent effects in our setup at 100 kHz and 40 kHz at this high average power, revealing different optimal fluence conditions for efficient conversion. The demonstrated sources with multi-100 mW average power at these high repetition rates combine high THz pulse energies and high repetition rate and is thus ideally suited for nonlinear THz spectroscopy experiments with significantly reduced measurement times. The presented result is a first benchmark for high average power THz time domain spectroscopy systems for nonlinear spectroscopy, driven by very high average power ultrafast Yb lasers.

Terahertz time domain spectroscopy (THz-TDS) is currently the cornerstone of many scientific breakthroughs. In particular, strong-field, few-cycle THz sources with electric fields >100 kV/cm have become ubiquitous to drive a variety of condensed matter systems and study their nonlinear response and dynamics [1–3]. To generate the corresponding broadband, high-energy, single-cycle THz pulses, the most commonly used approach is optical rectification in nonlinear materials exhibiting a $\chi^{(2)}$ response. The conversion efficiency in optical rectification scales with the driving pulse intensity, therefore high-energy laser systems in combination with large nonlinear crystals, such as lithium niobate ($LiNbO_3$) [4] or organic crystals [5] are typically used. As such, the main workhorse of nonlinear THz science remains Ti:Sapphire lasers, which provide mJ to J level femtosecond laser pulses from table-top systems. However, Ti:Sapphire lasers have well-known limitations in average power due to strong thermal aberrations in the gain medium, therefore the repetition rate of the driving pulse train is correspondingly low. Most commonly, reaching mJ pulse energies can thus only be achieved at repetition rates <5 kHz, and higher energies requires a corresponding reduction in repetition rate. The most recent THz energy record, for example, was achieved by Wu et al., generating 13.9 mJ of THz pulse energy at a repetition rate of 1 Hz and pump pulse energy of 1.2 J [4]. However, at such low repetition rates, measurement times are very long, due to the long waiting time to capture a single trace, and even longer times to average over many pulses. This, in turn, often results in a low dynamic range in the frequency domain, limiting many experiments. Single-shot techniques offer an alternative to mechanical delay scanning, which can retrieve the full THz time trace for each pulse in the pulse train and could reduce the measurement time [6]. However, although single-shot methods are currently progressing fast, these techniques usually suffer from comparatively low dynamic ranges, and scanning methods remain the method of choice for most spectroscopy experiments. In this regard, the combination of high THz pulse energy and a high repetition rate, i.e. a high THz average power is essential for reducing measurement times for experiments where multi-dimensional parameter studies are required.

With the advent of high average power, ultrafast ytterbium (Yb)-based laser systems, this compromise between repetition rate and pulse energy can be circumvented. These systems operate with a central wavelength around 1030 nm and typically have a pulse duration of a few hundred femtoseconds. Currently, off-the-shelf laser systems offer more than 100 W of average pump power, and up to 10 kW average powers have been demonstrated [7]. However, this new excitation regime demands the development of new guidelines for THz-TDS systems. In the last years, this has become an intense area of development, where several THz generation methods are currently being revisited with high average power excitation [8,9]. To name some of the most remarkable recent developments, in [10] Buldt et al. demonstrated 640 mW of average power at 500 kHz using a two-color plasma source in a gas-jet, driven by 16 coherently combined rod-type fiber laser delivering 633 W [10]. The corresponding source is an ultra-broadband THz source (>20 THz) with correspondingly moderate power spectral density.

For accessing even higher conversion efficiencies, and higher power spectral densities in the frequencies <2 THz, the tilted pulse front method in $LiNbO_3$ as introduced by Hebling et al. [11], is also a very attractive alternative for high average power excitation. In fact, $LiNbO_3$ has a high nonlinear coefficient, can be grown in large sizes, has low multi-photon absorption at common near-infrared wavelengths and generally has a high damage threshold, thus enabling high conversion efficiencies >1% [12] when velocity matching constraints are managed. Using this technique in combination with high average power lasers, Kramer et al.

demonstrated 144 mW using an Yb-based laser at 100 kHz repetition rate [13], and, at much higher repetition rates, Meyer et al. demonstrated 66 mW at 13.4 MHz [14]. An overview of the evolution of the average power of $LiNbO_3$ based THz sources is presented in Fig. 1. In this article, we demonstrate an increase of a factor of 4x compared to previously demonstrated tilted pulse front $LiNbO_3$ THz sources and achieve up to 643 mW of average power at 40 kHz repetition rate. The corresponding source has a THz pulse energy of 16.1 µJ, and reaches an estimated electric field of 340 kV/cm which can be further optimized to MV/cm. We also study the same source at a repetition rate of 100 kHz and achieve up to 445 mW of THz average power. We present performance at 40 kHz and 100 kHz and show quantifiable influence of thermal effects in this excitation regime. The achieved record high average power single-cycle source will be a very attractive driver for high-speed, nonlinear spectroscopy investigations.

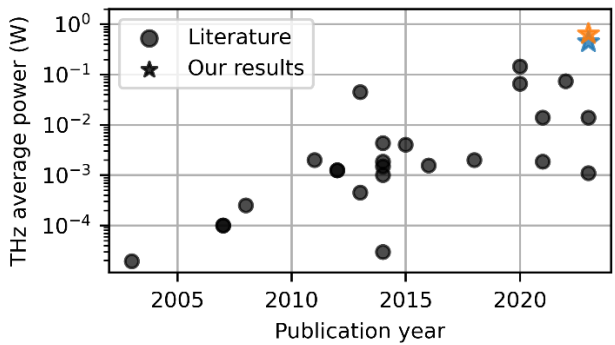

Fig. 1 Increase in THz average power of single-cycle THz sources based on the tilted-pulse front method in $LiNbO_3$ over the years with our result at 100 kHz (blue) and 40 kHz (orange).

## 1. Experimental setup

The experimental setup is presented in Fig. 2. A commercially available regenerative amplifier laser based on thin-disk technology is employed as the pump source (DIRA 500-10). The laser reaches up to 500 W of average power at three repetition rates (10 kHz, 40 kHz and 100 kHz), thus enabling us correspondingly to reach pulse energies of 50 mJ, 12.5 mJ and 5 mJ, at a pulse duration of approximately 750 fs. In this experiment, we limited ourselves to 100 kHz and 40 kHz, because of the limited size of the available $LiNbO_3$ crystals. The central wavelength of the laser is 1030 nm. The setup uses all high-reflective dielectric mirrors with a diameter of 50.8 mm (Optoman, ULLM5SHL) to avoid thermal effects and damage. A 45° output coupler (OC) is used to divide the beam into a pump beam for generation and a probe beam for electro-optic sampling (EOS) with a ratio of 0.2%. The pump beam is guided through a $\lambda/2$-waveplate (WP) and a thin-film polarizer (TFP). As the WP is mounted in a motorized rotation mount, the combination of WP and TFP allows for the repeatable and precise selection of pump power reaching the crystal. The reflection of the TFP is guided towards a fused silica transmission grating (LightSmyth/Coherent, 1600 lines/mm) in order to tilt the pulse front. Subsequently, the beam is imaged from the grating into the crystal with two 50.8 mm diameter lenses with 300 mm and 200 mm focal length, respectively.

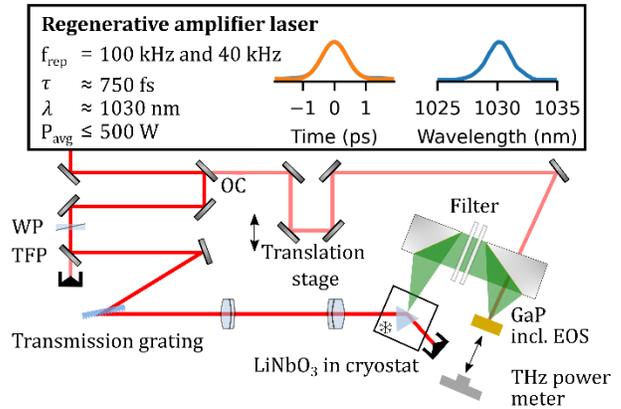

Fig. 2 Schematic of the experimental setup. The inset shows the autocorrelation trace and spectrum of the laser at 100 kHz with similar specifications at 40 kHz. The pump beam (strong red) and probe beam (weak red) are split by an output coupler (OC). The resulting THz (green) is detected by gallium phosphide (GaP) incl. EOS or a THz power meter.

Prior to reaching the crystal, the beam passes through an anti-reflective (AR) coated window of the cryostat (Janis/LakeShore, ST100). For this experiment, the temperature is maintained at 80 K to reduce the absorption coefficient of $LiNbO_3$ at THz frequencies. The pump beam enters the AR-coated crystal (Oxide, stoichiometric $LiNbO_3$ with an input area of 20 mm × 20 mm) within the cryostat. Due to the triangular geometry of the crystal and the pulse front tilt of the pump pulse, the velocity matching condition is fulfilled, allowing for the efficient generation of THz radiation over a significant crystal length. The pump undergoes total internal reflection and is coupled out on the right side of the prism. The THz pulse, however, travels perpendicularly out of the crystal and is collected by an off-axis parabolic (OAP) mirror. A stack of filters, consisting of high-density polyethylene (HDPE), is employed to suppress residual scattered pump light and parasitic second harmonic generated (SHG) light. The THz pulse is either focused onto a gallium phosphide (GaP, 0.5 mm) crystal for EOS or, while blocking the probe beam, onto a THz power meter (Ophir, 3A-P-THz). In addition to the motorized rotation stage, all relevant degrees of freedom are motorized in the setup, which allows for automated, multi-dimensional parameter scans. The position and angle of the final mirror prior to the grating are also motorized, as are the angle of the grating, the position of both lenses along the beam's propagation direction, and the position of the cryostat. The setup itself was generally designed following the guidelines presented in Kroh et al. [15], which were introduced to study the parameter sensitivities in a tilted pulse front setup with a low-repetition rate laser.

## 2. Results and discussion

The obtained THz average power for various pump powers at the two explored repetition rates is presented in Fig. 3 a). We start our exploration with the highest available repetition rate of 100 kHz. The collimated beam diameter before the grating was adjusted to achieve peak efficiency at maximum available pump power. The beam diameter was approximately 7.3 mm ($1/e^2$) at the crystal, and only slightly increased to 7.5 mm ($1/e^2$) in the case of 40 kHz. The expansion of the beam diameter resulted in a slight increase in clipping on the grating, which in turn led to a reduction in the available pump power on the crystal.

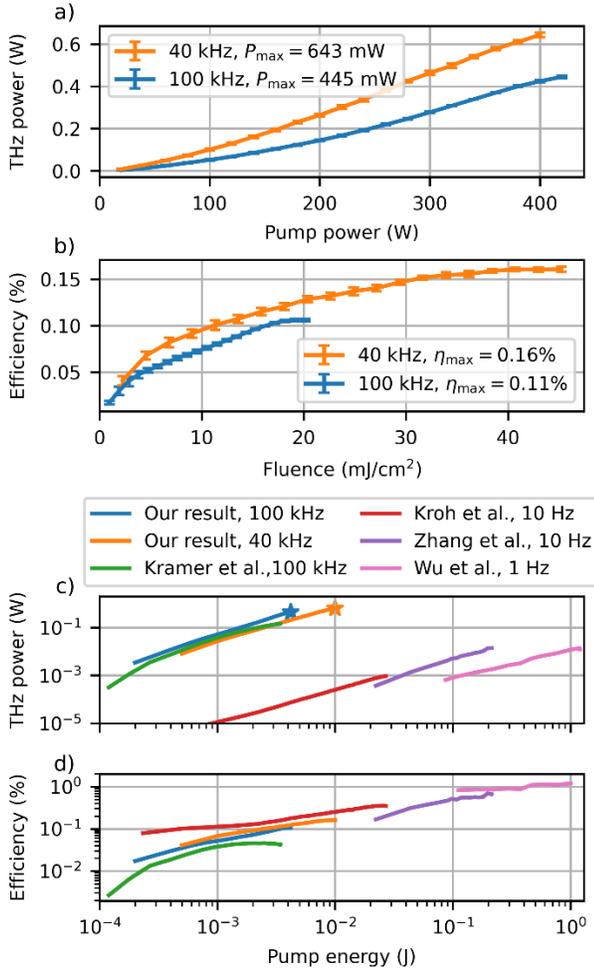

Fig. 3 a) Obtained THz power at 40 kHz and 100 kHz for various pump powers. b) The beam diameter and fluence were selected as such, that the shown efficiency peaks at maximum pump power. The error bars indicate the standard deviation of 10 repetitions for these power curves. Comparison to other results in terms of c) average power and d) efficiency from [4,13,15,16] with the respective repetition rate included in the legend.

At the repetition rate of 100 kHz, a max. THz average power of 445 mW was achieved. By reducing the repetition rate and having nearly the same pump power available, the pump pulse energy was increased. Reducing the repetition rate to 40 kHz resulted in the highest single-cycle THz average power of 643 mW from any nonlinear medium pumped by a table-top laser system. The previous record for lithium niobate was 144 mW at room temperature and 100 kHz of repetition rate [13].

Fig. 3 b) shows the achieved efficiency in terms of fluence, showing that in both repetition rates, the optimal conversion efficiency before rollover is reached is significantly different. This illustrates the difficulty of scaling the average power in general using nonlinear crystals: for a higher repetition rate, the onset of thermal effects occurs at lower fluences, limiting the applicable intensity before observing a rollover in conversion efficiency. In fact, one notable distinction between this setup and those with high pulse energies but low repetition rates previously reported in the literature is the optimal fluence achievable in absolute terms. While in systems with low pump average powers, fluences above 230 mJ/cm$^2$ at cryogenic temperatures were achievable prior to saturation [15], for the high-power system presented here at 100 kHz, the optimal fluence was 20 mJ/cm$^2$, corresponding to approximately 25 GW/cm$^2$ peak intensity, at the maximum pump power. At 40 kHz, these values were improved to 45 mJ/cm$^2$ and 57 GW/cm$^2$. The differences between our system in terms of reachable efficiency and other systems in the literature at lower repetition rates are presented in Fig. 3 c) and 3 d). Whereas absolute values in very different experimental conditions are difficult to directly compare, the trend is clearly visible: reaching high repetition rates comes at a sacrificed conversion efficiency. Revealing the exact mechanism for these observations requires a more detailed investigation of the fundamental material property changes in $LiNbO_3$, however the influence of thermal effects can certainly not be neglected and need to be studied in more detail in future works. Possibly, a better cooling geometry could slightly improve these results; however, heat extraction is ultimately limited by the limited thermal conductivity of the material.

After characterizing the THz average power, the power meter was replaced by a GaP crystal and an EOS setup. Fig. 4 a) shows the averaged THz traces in the time domain for various pump powers. In addition to the single-cycle nature of the THz pulse, the expected increase in signal strength can be observed by increasing the pump power. Due to the long pulse duration of the laser, the bandwidth is not limited in the first place by the phonon resonances in $LiNbO_3$ but rather by the comparatively narrow bandwidth of the pump pulses. Moreover, the readout pulse has the same pulse duration as the pump, thereby restricting the bandwidth of the THz pulse. By blocking the pump power but maintaining all other settings, a "dark" trace is captured, providing insight into the frequency-dependent noise profile of the THz-TDS system. The dark trace facilitates the visualization of the bandwidth of $\approx 1.5$ THz and provides insight into the peak dynamic range of $\approx 60$ dB, as it can be observed in Fig. 4 b). The moderate dynamic range for the presented source (relatively to the available average power) is attributed to the non-ideal imaging conditions in this first realization aiming at power optimization rather than application of the THz-TDS. The inset of Fig. 4 b) provides evidence of the non-ideal imaging condition. The configuration of the OAPs depicted in Fig. 2 is optimal for a point source, which is only poorly valid in this case because the pump beam and the resulting THz beam is relatively large. Recent simulations from Chopra & Lloyd-Hughes demonstrated that the alignment of OAPs is crucial for maintaining optimal focusing, as any deviations from the correct alignment can result in a larger, distorted THz beam and reduced electric field [17]. Initial measurements at 40 W of pump power with a THz camera showed a beam diameter of close to 3 mm (1/e$^2$), which would give an electric field of $\approx 340$ kV/cm at maximum power, using the guidelines established by [18]. We note however, that it is challenging to accurately estimate the electric field due to the non-ideal imaging conditions and the rather long EOS probe pulses, and these values should be only used as indicative. A future improved imaging setup will take these aspects into account to achieve high electric field values. We note in Fig. 4 a) a ps scale, power dependent time-shift which will be critical to harness for future spectroscopy experiments, where this could be wrongly attributed to a phase shift caused by an eventual sample. In addition to the aforementioned geometrical arguments, thermal effects (for example via the dn/dT of the material) due to both residual linear and nonlinear absorption could also explain this behavior.

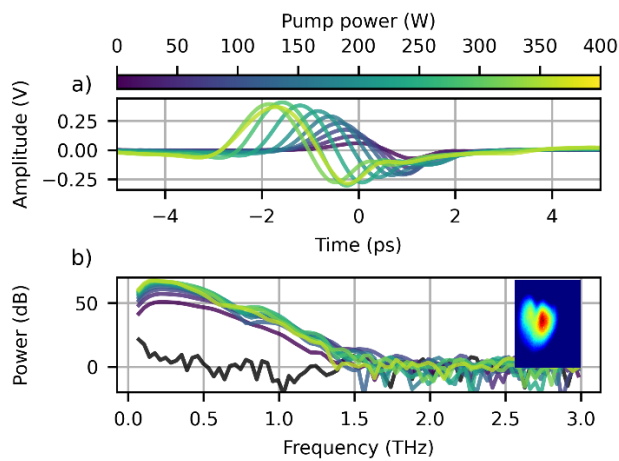

Fig. 4 a) THz traces obtained by EOS for various pump powers at 40 kHz. The single-cycle nature can be seen as well as a power dependent time shift of the trace. b) Corresponding power spectra, where the black trace shows the noise floor of the system when the THz beam is blocked. Inset: THz beam profile at 100 kHz and 40 W, indicating an average beam diameter of approximately 3 mm ($1/e^2$).

The moderate thermal conductivity of $0.3\,\text{W}/(\text{K}\cdot\text{cm})$ of $LiNbO_3$ at 80 K [19] and the Gaussian beam profile result in a gradient within the crystal, with only the top surface in contact with the copper cooling finger of the cryostat. Short experimental durations, as conducted for the power curves in Fig. 3 a), were possible but brought the cryostat to the limit of cooling the sample. Another possible limitation at high average power is increased heat due to two-photon absorption of 515 nm and potential residual photorefractive effect [20]. The crystal is doped with 1.3 mol% of magnesium which reduces photorefractive effects, however these can still be presence at the large average power in place. Given the high average intensity used, the effect could lead to refractive index modulation, which in turn has further effects on the optimal tilting angle, shifting the effective interaction volume of the pump and THz beam. Combined with imaging of the OAPs, this could explain the phase shift in the time domain. In a future comprehensive study, we will investigate in detail the underlying mechanism of the observed phase shifts and thermal effects.

## 3. Conclusion and outlook

We demonstrate the highest THz average power single-cycle source from a table-top laser system with 643 mW and 16.1 µJ of energy at a high repetition rate of 40 kHz. At a higher repetition rate of 100 kHz, we also achieve high average power of 445 mW. Additional explorations and optimization of the maximum electric field in this regime was prevented by damage on the available crystal. We note that the damage of the crystal was not due to the laser intensity but rather by an accidental misalignment caused by the automatic alignment used. In continuation experiments, and according to the findings of this paper, a higher pump pulse energy with a larger pump spot at the same average power should result in higher conversion efficiency. By setting the laser to a 10 kHz repetition rate, up to 50 mJ of pulse energy is available. However, at the time of the experiment, we did not have sufficiently large crystals to perform this experiment. With larger crystals (30 mm × 30 mm) we expect a watt-level source with electric fields exceeding 1 MV/cm at 10 kHz repetition rate to be achievable in the near future. In the meantime, the current source is already an extremely promising starting point for nonlinear spectroscopy experiments with only small modifications in the cooling geometry and imaging.

**Funding.** XXX

**Acknowledgments.** We thank Alan Omar, Kore Hasse, and Frank Wulf for fruitful discussions. This work was supported in part by DFG of the SFB/TRR196 MARIE project M01 and C07 and the DFG project PR1413/3–2, in part by the DFG under Germany's Excellence Strategy— EXC-2033— Projektnummer 390677874—Resolv, and in part by project "terahertz.NRW" program "Netzwerke 2021," an initiative of the Ministry of Culture and Science of the State of Northrhine Westphalia.

**Disclosures.** Part of the results were presented at the conferences "Infrared, Millimeter and Terahertz Waves (IRMMW-THz)" and "Conference on Lasers and Electro-Optics (CLEO)" in USA and Europe.

**Data availability.** Data underlying the results presented in this paper are available in [21].